\documentclass[prl,reprint,showpacs,nofootinbib]{revtex4-1}

\usepackage{amsmath,empheq}
\usepackage{amsfonts}
\usepackage{amsthm}
\usepackage{amssymb}
\usepackage{graphicx}
\usepackage{hyperref}
\usepackage{soul}

	\newcommand{\ncd}{\newcommand}
	\ncd{\mrm}    {\mathrm}
	\ncd{\beq} {\begin{equation}}
	\ncd{\eeq} {\end{equation}}

	\def\d{{\rm d}}

\begin{document}
	\title{Maximally Symmetric Spacetimes emerging from thermodynamic fluctuations}
	\date{\today}

	\author{A. Bravetti}
        \email{bravetti@correo.nucleares.unam.mx}
	\affiliation{Instituto de Ciencias Nucleares, Universidad Nacional Aut\'onoma de M\'exico,\\ AP 70543, M\'exico, DF 04510, Mexico.}

	\author{C. S. Lopez-Monsalvo}
        \email{cesar.slm@correo.nucleares.unam.mx}
	\affiliation{Instituto de Ciencias Nucleares, Universidad Nacional Aut\'onoma de M\'exico,\\ AP 70543, M\'exico, DF 04510, Mexico.}

	\author{H. Quevedo}
        \email{quevedo@nucleares.unam.mx}
	\affiliation{Instituto de Ciencias Nucleares, Universidad Nacional Aut\'onoma de M\'exico,\\ AP 70543, M\'exico, DF 04510, Mexico.}

	\begin{abstract}
In this work we prove that the maximally symmetric vacuum solutions of General Relativity emerge from the geometric structure of statistical mechanics and thermodynamic fluctuation theory. To present our argument, we begin by showing that the pseudo-Riemannian structure of the Thermodynamic Phase Space is a solution to the vacuum Einstein-Gauss-Bonnet theory of gravity with a cosmological constant. Then, we use the geometry of equilibrium thermodynamics to demonstrate that the maximally symmetric vacuum solutions of Einstein's Field Equations -- Minkowski, de-Sitter and Anti-de-Sitter spacetimes -- correspond to thermodynamic fluctuations. Moreover, we argue that these might be the only possible solutions that can be derived in this manner. Thus, the results presented here are the first concrete examples of spacetimes  effectively emerging  from the thermodynamic limit over an unspecified microscopic theory without any further assumptions.
	\end{abstract}


\maketitle

The centennial Einstein's Field Equations of General Relativity \cite{Einstein1915}
establish a direct link between gravity and geometry. 
{It is easy to see that these equations 
are highly non-trivial  even in the absence of a `material' source. 
The quest for finding solutions to the vacuum Field Equations started even before their publication in final form. 
Examples of these are the class of `black-hole' solutions, 
and de-Sitter (dS) and Anti-de-Sitter (AdS) spacetimes in the case of a non-vanishing cosmological constant.}   
In addition to Einstein's dynamical equations, 
the celebrated Bekenstein-Hawking formula for the entropy of a black-hole \cite{Bek}
has set up the relationship between spacetime geometry and thermodynamics,
which was gracefully confirmed by {the calculation of the temperature of a horizon based on quantum field theory} of
Gibbons and Hawking  \cite{Hawking,GH}.
Although such derivation is limited to a semi-classical approximation, it shows that the thermodynamic properties of black holes satisfy relations that are completely analogous to the Laws of thermodynamics \cite{BHsLaws}, which is generally believed not to be a mere coincidence.

In {our view}, we are now in a situation similar to that of Clausius and Carnot in the mid XIX century. When they formulated the macroscopic laws of thermodynamics  
they had no \emph{a priori} knowledge of the microscopic properties of gases. Nevertheless, they found a set of 
general relations among macroscopic variables which are independent of the {microphysics.} 
It was only later -- through the works of Boltzmann and Gibbs in particular -- that it was {understood} that the macroscopic Laws of thermodynamics {admit a} formulation
in terms of the statistical mechanics of systems with a large number of degrees of freedom.
Remarkably, it was found that some properties of the thermodynamic behavior of large systems -- e.g. concepts like entropy, temperature and phase transitions
 -- \emph{emerge} as a result of the thermodynamic limit.
This means that systems with a large number of degrees of freedom {exhibit} a \emph{qualitatively new} behaviour with respect to those obeying the same microscopic
laws of motion but having a small number of degrees of freedom.

Returning to gravity,  at the present time we {have} only {gained some knowledge about}  the laws of the macroscopic thermal behavior of gravitational vacuum solutions, 
but we {completely} lack a microscopic description.
Moreover, the puzzle has acquired more ingredients.
It has been found that in many cases the thermodynamics of black holes essentially resembles that of standard thermodynamic systems, in particular in the presence
of a non-vanishing cosmological constant {which is considered as an intensive thermodynamic variable 
 (see e.g. \cite{MannGalaxies,Dolan} and references therein).}
Another connection between thermodynamics and gravity is that it was proved that Einstein's equations 
can be rewritten in the form of the First Law of thermodynamics \cite{Jacobson,Padmanab}. 
Finally, gravity and thermodynamics share a fundamental property: both theories are \emph{universal}. That is, they make general statements valid for any systems, 
independently of their
internal description.

All these facts have motivated the idea that gravity {should not} be considered as a fundamental {interaction} but, instead, as an \emph{emergent phenomenon}.
{Two crucial and related questions are therefore}
what does it mean for gravity and spacetime to be {emergent}. 
{On the one hand, the former question has been recently addressed in Verlinde's proposal of gravity as an entropic force \cite{Verlinde}. 
Such approach is strongly based on the \emph{holographic principle}, as formulated by 't Hooft and Susskind \cite{tHooft,Susskind} and the \emph{thermodynamic limit}. 
On the other hand, regarding the emergence of space, Dieks et-al \cite{PhilosophyOfEmergence} pointed out that   
`if space is emergent, it must be the thermodynamic limit that does the work'. In this perspective
the thermodynamic limit is the only ingredient that is really needed to give a precise account of emergence
and in particular Verlinde's proposal does not seem to realize a real emergence of spacetime \cite{PhilosophyOfEmergence}.

However, emergence of spacetime itself is a crucial point in the whole picture of emergent models of gravity, since it may resolve automatically many of the challenges
such as e.g. recovering Lorentz invariance and the relativity principle \cite{Carlip}.
In this sense, Zhao proposed recently a suggestive model in which spacetime itself is emergent, based on the geometry of thermodynamics and showed that in this context
not only the principle of relativity arises naturally, but it can be given a thermodynamic interpretation \cite{Zhao1}.
In fact, a related instance of emergence stems from the geometrisation of thermodynamics, as proposed originally by Rao, Weinhold, Ruppeiner, Hermann and Mrugala  
(see \cite{amari,wein1975,rupp1979,rupp1995,Hermann,mrugala1,mrugala2,SalamonBerryPRL,Schlogl,BrodyRivier,Crooks,CrooksPRL2012}). 
From this perspective,
thermodynamic fluctuations of the macroscopic observables define a metric tensor on an abstract  space, whose components in the natural thermodynamic 
coordinates are given by
	\beq\label{metriconE}
	g _{ab} =\langle \left(F_a - \langle F_a \rangle \right) \left( F_b - \langle F_b \rangle \right) \rangle = \left[{\rm Var}(\d h)\right]_{ab}, 
	\eeq
where the $\{F_a\}$ is the set of $n$ oservables \cite{mrugala1990,TPSSASAKI,CONTACTHAM}.
Here, the average is computed with respect to Gibbs' probability distribution $\rho$, with $h = - \ln \rho$ the microscopic entropy and `Var' stands for the variance.

In this work, motivated by the individual connections between geometry, gravity and thermodynamics depicted in Figure \ref{fig1},
we propose a new scheme for the emergence of spacetime, which follows the right part of the triangle in Figure \ref{fig1}.
That is, starting from the thermodynamic limit and without resorting to the holographic principle, we derive a geometry of fluctuations which we identify with spacetime 
through Einstein's equations.
Thus, let us agree on three assumptions: 
(i) There exists a microscopic theory underlying the gravitational phenomena, 
(ii) the microscopic theory obeys 
Boltzmann-Gibbs'  statistical mechanics, and 
(iii) spacetime itself is emergent and gets identified with a \emph{Legendre} sub-manifold representing a thermodynamic system.
The first two assumptions seem natural from a standard thermodynamic perspective, while the third one is motivated by the results
of Zhao \cite{Zhao1}.

Using solely the assumptions stated in the previous paragraph and the geometric construction in \cite{mrugala1990,TPSSASAKI,CONTACTHAM},
we prove the following results: (1) The (unique) metric structure of the Thermodynamic Phase Space (TPS), resulting from the maximum entropy principle, 
 is a vacuum solution to the Einstein-Gauss-Bonnet {theory of gravity with a positive
cosmological constant in any odd dimensions, and (2) imposing the First Law of thermodynamics as the condition that first order fluctuations vanish, we obtain that 
the Minkowski, {dS and AdS spacetimes in any dimensions} naturally emerge from second order fluctuations.

\begin{figure}
\begin{center}
\includegraphics[width=1\columnwidth]{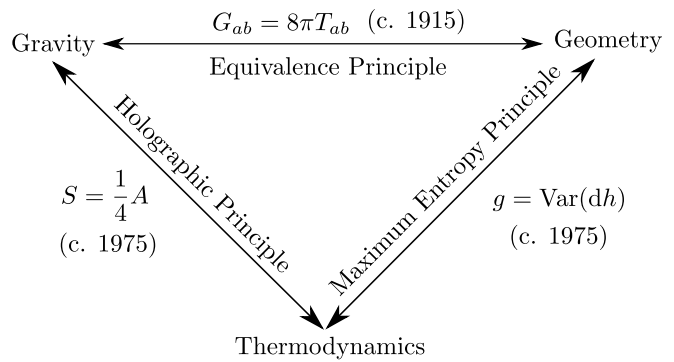}
\caption{A diagram motivating our approach. The top line shows Einstein's field equations and highlights the way they establish a relationship between 
gravity and geometry, which was motivated by the equivalence principle. The left line shows Bekenstein-Hawking's formula for the entropy of horizons in General Relativity,
which establishes a relationship between thermodynamics and gravity and has been motivated \emph{a posteriori} by the holographic principle. 
The right hand side of the diagram shows the geometry of thermodynamic fluctuations, which curiously was derived by Weinhold, Ruppeiner, Mrugala and others along the same years as the 
Bekenstein-Hawking derivation of the area law. 
This formulation establishes the missing link between thermodynamics and geometry and can be motivated through the maximum entropy principle.
}
\label{fig1}
\end{center}
\end{figure}

	

In \cite{mrugala1990} it was shown that 
given \emph{any} microscopic theory 
in which the average values of a given set of fluctuating observables -- $\langle F_{a}\rangle=p_{a}$, with $a=1,\dots,n$  --  
with respect to a thermal (Gibbs) distribution 
	\beq
	\rho={\rm exp}(-w-F_{a}q^{a})
	\eeq
are known, one can use Boltzmann-Gibbs' statistical mechanics to construct a macroscopic phase space -- the Thermodynamic Phase Space (TPS) --
which is a $(2n+1)$-dimensional manifold whose coordinates are the mean values $p_{a}$, the corresponding Lagrange multipliers $q^{a}$
and the Lagrange multiplier $w$ related to the normalization of $\rho$.
Moreover, it was shown (c.f. \cite{TPSSASAKI,CONTACTHAM}) that the first moment of the differential of the entropy $h=-{\rm ln}(\rho)$ 
naturally endows the TPS with a contact $1$-form
	\beq
	\eta=\langle\d h\rangle=\d w+p_{a}\d q^{a}.
	\eeq
The kernel of such $1$-form defines the First Law of thermodynamics for quasi-static processes,
and therefore its maximal integral sub-manifolds -- the Legendre sub-manifolds -- are \emph{equilibrium manifolds} 
for any thermodynamic system.
The second moment of the differential of the entropy defines a pseudo-Riemannian structure on the TPS, whose line element is given by
	\beq\label{G}
	\d s^{2}_{\rm TPS}=\langle \left(\d h\right)^{2}\rangle=\langle \d h \rangle^{2}+{\rm Var}(\d h)=\eta^{2} - \d p_{a}\d q^{a},
	\eeq
where ${\rm Var}(\d h)$ is the variance. 
Notice that the term $\eta^{2}$ arises from the square of the first order fluctuations, while $\d p_{a}\d q^{a}$
takes account of second order fluctuations.
 The signature of \eqref{G} is $(n+1,n)$
and it has peculiar geometric properties, that is, it is a para-Sasakian and $\eta$-Einstein manifold which is locally isomorphic to 
the hyperbolic Heisenberg group \cite{TPSSASAKI}.

As anticipated, the equilibrium manifold of any thermodynamic system can then be given a precise geometric definition as a Legendre sub-manifold of the equation
$\eta=0$.
It turns out that such sub-manifolds have at most $n$ degrees of freedom and that the induced distance
is given by
	\beq\label{g}
	\d s^{2}_{\rm Eq}=-\frac{\partial^{2} w}{\partial q^{a} \partial q^{b}}\d q^{a}\d q^{b},
	\eeq
with $w={\rm ln}(\mathcal Z)$ being the thermodynamic potential and $\mathcal Z$ the partition function corresponding to the proper ensemble.
Eq. \eqref{g} is the thermodynamic version of the Fisher-Rao distance, which measures the thermodynamic fluctuations out of the equilibrium values 
of the macroscopic observables \cite{amari,Schlogl,BrodyRivier,Crooks,CrooksPRL2012}.
Notice also that the construction of the TPS with its metric structure and that of the integral manifolds of equilibrium states with the induced metric
are \emph{independent} of the detailed description of the microscopic theory, provided this has a thermal (Gibbs) equilibrium distribution.
The {crucial point} is that the geometric nature of the TPS and of  its sub-manifolds emerge only from the thermodynamic limit and are \emph{indifferent} of the microscopic details.
Notice also that the TPS is $(2n+1)$-dimensional, while the equilibrium manifold is $n$-dimensional.

In {this context}, Zhao has proven that the First Law of thermodynamics and the equations of state, are \emph{invariant}
under the group of general diffeomorphisms of the ($n$-dimensional) equilibrium manifold and argued 
that this can explain the origin of the principle of relativity in an emergent picture of spacetime.
Furthermore, \emph{assuming}  that the signature of the metric on the equilibrium
manifold must be Lorentzian,  Zhao has also formulated 
an invariant version of the Second Law of thermodynamics \cite{Zhao1}.
A related derivation was obtained in \cite{qsv15} in the context of the geometrothermodynamics programme \cite{quev07}. 
In our case we will depart from Zhao in two main aspects. On one side, we identify the equilibrium manifold exactly with spacetime, while Zhao suggests that there might be 
more degrees of freedom related to the volume and the number of particles. In this way, the general group of diffeomorphisms of the equilibrium manifold
directly coincides with the group of diffeomorphisms of spacetime and the principle of relativity is automatically recovered, without the ambiguities related to variations
of the volume or number of particles encountered in \cite{Zhao1}.
On the other side,  the metric structure on the TPS is fixed by the statistical derivation  -- c.f. \eqref{G} -- and therefore the form of the metric on the equilibrium manifold is given by \eqref{g}.
Here, the signature of the metric is \emph{determined} by the convexity conditions implied by the Second Law. Indeed, all that we know is that \eqref{g} is the Hessian of some thermodynamic potential,
depending on the ensemble. Therefore it is known from standard thermodynamics that the Second Law implies e.g. that the Hessian of entropy must be negative definite (i.e. Euclidean), while 
the Hessian of its partial transform $-\beta F=S-E/T$ has a Lorentzian signature (see also \cite{thermometric} for a related discussion). 
To conclude, we will assume in the following that the ensemble considered is the canonical ensemble and
therefore $w=-\beta F$.


The simplest extension of General Relativity which preserves the order of the differential equations determining the spacetime metric  is Einstein-Gauss-Bonnet gravity (EGB). 
Such an extension yields non-trivial dynamics in any spacetime dimension greater than four, where the GB correction 
is  a topological  term in the action.} Furthermore, it has been also shown that  in the low energy limit of string theories the leading quadratic curvature correction
to the Einstein-Hilbert action is precisely the EGB term \cite{stringEGB}. 
Since the TPS is endowed with a metric structure reminiscent of an Einstein manifold (in fact, an $\eta$-Einstein manifold), it is natural to wonder whether this is 
a solution of EGB vacuum field equations.
It turns out that in any odd-dimension $D=2n+1$,  substituting the metric \eqref{G} into the EGB vacuum field equations  with a cosmological constant
yields the algebraic system
	\begin{empheq}[left=\empheqlbrace]{align}
	\label{z1}
	& \Lambda-30n(n-1)\alpha-3n \,\,= 0,\\
	\label{z2}
	& \Lambda-6(n-1)(n-4)\alpha-n+2 \,\,=  0,
	\end{empheq}
whose only solution is $\Lambda=n/2$ and  $\alpha=-1/[12(n-1)]$.

Therefore, the TPS metric \eqref{G} \emph{is a solution to the vacuum EGB field equations in any odd-dimension $D$ with a cosmological constant $\Lambda=n/2$ and
with a GB parameter $\alpha=-1/[12(n-1)]$.}
This is the first result of the present work.

Notice that this solution is the \emph{unique} geometry of the TPS arising from the maximum entropy principle and that we have made no  
assumption here other than Boltzmann-Gibbs' statistical mechanics.
It is therefore an interesting fact that a solution to the vacuum field equations -- which are highly nontrivial  -- arises from such basic assumptions. 
We remark that the distance \eqref{G} in the TPS is composed by two terms: one is the square of the linear fluctuations, which should be zero at exact thermodynamic
equilibrium -- but it is not vanishing here, meaning that we are only {\it near} thermodynamic equilibrium -- 
and the other term is given by second order fluctuations. 
The fact that this solution is not exactly at thermodynamic equilibrium is mirrored by the fact that
the spacetime described by \eqref{G} has $(n+1)^{2}< D(D+1)/2$ Killing vector fields and therefore it is not maximally symmetric \cite{TPSSASAKI}.
 We will see  that, by imposing exact thermodynamic equilibrium, one is left with maximally symmetric spacetimes in $d=n$ dimensions.
 
Finally, let us note that the coordinates in which the solution \eqref{G} 
has been derived  have a direct thermodynamic meaning, however, their interpretation as spacetime coordinates is not \emph{a priori} clear. 
A further physical study of this spacetime solution is therefore needed. A striking feature that should be mentioned is that such spacetime 
apparently contains information reminiscent 
of quantum relations, since it is locally isomorphic to the Heisenberg group \cite{TPSSASAKI}.

\vspace{.2cm}


Now given the discussion about the emergent nature of gravity and in particular of spacetime,
one would be pleased to find spacetime manifolds which emerge as spaces of exact equilibrium states for given systems.
In the following 
we show that the Minkowski, dS and AdS spacetimes naturally emerge as macroscopic systems at equilibrium.   
We also give some reasons that hint to the fact that these are the only vacuum solutions (possibly including the cosmological constant) at 
absolute thermodynamic equilibrium within Einstein's theory in four dimensions.
To do so, we consider the equilibrium submanifolds  
 -- which by construction are $n$-dimensional -- equipped with their induced distance \eqref{g}.  

From equation \eqref{g} we notice that all equilibrium geometries have a very peculiar property, that is, they are \emph{Hessian metrics} \cite{ShimaBook}. 
Our third assumption implies that the metric \eqref{g} must also satisfy Einstein's field equations. In particular, we will consider vacuum equations and thus \eqref{g}
must be an \emph{Einstein metric} \cite{EinsteinMetrics}.
Therefore the class of geometries for the equilibrium manifold gets very constrained.
Indeed, from all the known examples in the geometry of Hessian manifolds \cite{ShimaBook}, we find that only two families of thermodynamic potentials generate Hessian
metrics which are also Einstein. These are 
	\beq\label{Mink}
	w(q^{a})=\sum_{i=1}^{n}f_{i}(q^{i})
	\eeq
and
	\beq\label{dSAdS}
	w(q^{a})=\pm C^{2}\,{\rm ln}\left(q^{1}+\sum_{i=2}^{n}k_{i}(q^{i})\right),
	\eeq
where the $f_{i}$'s and $k_{i}$'s are free functions of the corresponding variable $q^{i}$, the only restriction being that the metric is non-degenerate, which implies
the $k_{i}$'s to be non-linear functions.
One can directly verify that all the geometries generated by \eqref{Mink}
have vanishing Riemann tensor, and therefore the only spacetime thus generated is the flat Minkowski spacetime.
Moreover, all the functions in the family \eqref{dSAdS} 
generate only two spacetimes:  {dS}$_{n}$ and  {AdS}$_{n}$, which correspond to the $\pm$ signs in \eqref{dSAdS}, respectively.
Here, the constant $C^{2}$ is directly related to the radius $l$ of dS$_{n}$ (resp. AdS$_{n}$)  by the relation $C^{2}=l^{2}/4$.
Therefore, we have proved the following result:
\emph{the three possible maximally symmetric  solutions of the vacuum Einstein's field equations, 
Minkowski, de-Sitter and Anti-de-Sitter, emerge as the geometry of the equilibrium manifold of some macroscopic system
obtained by taking the thermodynamic limit over an 
unspecified microscopic theory.
}


Now that we have found the spacetimes of Einstein's theory emerging from the thermodynamic limit, it remains to 
relate the `thermodynamic' coordinates $q^{a}$  
 to a more standard set of expressions for spacetime manifolds. 
To do so,  let us consider four dimensional de-Sitter in the `static' patch, 
i.e.
	\beq\label{dSspheric}
	ds^{2}_{\rm dS}=-\left(1-\frac{r^{2}}{l^{2}}\right)\d t^{2}+\left(1-\frac{r^{2}}{l^{2}}\right)^{-1}\d r^{2}+r^{2}\d \Omega^{2},
	\eeq
where $\d\Omega^{2}=\d \theta^{2}+{\rm sin(\theta)^{2}\d \varphi^{2}}$, and find the explicit diffeomorphism. 
{Choosing a quadratic expression for the functions $k_{i}$ in \eqref{dSAdS}}, we obtain
	\beq\label{dSAdS2}
	w(q^{a})=\frac{l^{2}}{4}\,{\rm ln}\left(q^{1}+\frac{1}{2}\sum_{i=2}^{4}(q^{i})^{2}\right)
	\eeq 
for the potential and we find that the sought isometry is
	\begin{align}
	\label{diffeo1}
	q^{1} & =\frac{\left(2r^{2}-l^{2}\right){\rm e}^{2t/l}}{2\left(r^{2}-l^{2}\right)}, \quad 
	q^{2}=\frac{r{\rm cos}\theta{\rm e}^{t/l}}{\sqrt{r^{2}-l^{2}}},\\
	\label{diffeo3}
	q^{3} & =\frac{r{\rm sin}\theta{\rm sin}\varphi{\rm e}^{t/l}}{\sqrt{r^{2}-l^{2}}}, \quad 
	q^{4}=-\frac{r{\rm sin}\theta{\rm cos}\varphi{\rm e}^{t/l}}{\sqrt{r^{2}-l^{2}}}.
	\end{align}

Given the explicit expression for the change from \emph{thermodynamic} to \emph{spacetime} coordinates, we can read out the physical significance, if any, 
of  the thermodynamic potential \eqref{dSAdS2}. 
Consider the interior region $r< l$ of \eqref{dSspheric}. The potential \eqref{dSAdS2} takes the form
	\beq\label{TDP}
	w=-\frac{l^{2}}{4}{\rm ln}\left[\left(\frac{r^{2}-l^{2}}{l^{2}}\right){\rm e}^{\frac{2t}{l}}\right],
	\eeq
whereas the relativistic analogue of the Newtonian potential is given by  \cite{Verlinde}
	\beq\label{NewtonP}
	\Phi=\frac{1}{2}{\rm ln}(-\xi_{a}\xi^{a})=\frac{1}{2}{\rm ln}\left[\left(\frac{l^{2}-r^{2}}{l^{4}}\right){\rm e}^{\frac{2t}{l}}\right]\,,
	\eeq
where $\xi$ is a timelike boost generator and  we have dropped an additive constant.
Thus, rescaling the Killing vector $\xi$ to $\tilde\xi=l\xi$ and computing the 
Newtonian potential with this new Killing vector $\tilde\xi$, it follows that
	\beq\label{wandPhi}
	{\Delta w}=-\frac{l^{2}}{2}\Delta\Phi\,.
	\eeq
{A similar expression was derived by Verlinde in \cite{Verlinde} but with much more stringent assumptions (c.f. \cite{Visser}) and it was used to motivate that gravity might have an entropic origin.}
This result is striking in our perspective, because here we have made no assumptions other than the thermodynamic nature of the system. 

 Notice that we could have used any other function in the family \eqref{dSAdS}. We decided to use this particular form because of its simplicity.
With a different expression for \eqref{dSAdS2}, the diffeomorphism would have been 
different, but the final value of the potential function would be unchanged. 

{Additionally}, we can obtain another interesting interpretation using \eqref{dSAdS2}. Recall that by definition $w$ is the logarithm of the partition function.
Therefore we can {read out} the partition function for {dS$_{4}$} to be 
	\beq\label{partitiondS}
	\mathcal Z=\left(q^{1}+\frac{1}{2}\sum_{i=2}^{4}(q^{i})^{2}\right)^{\frac{l^{2}}{4}}.
	\eeq
This expression looks like the partition function of a non-interacting system $\mathcal Z={\zeta}^{N}$. Provided this identification is correct, the number
of the degrees of freedom is proportional to $N=l^{2}/4$, i.e. to the area of the cosmological horizon.
Finally, considering that $w=-\beta F=S-E/T$, and using standard statistical mechanics applied to the partition function for spacetime,
we {find} out that $w=iI[g_{0},\Phi_{0}]$, {i.e. $w$} gets identified with the classical action for General Relativity (c.f. equation (6) in \cite{Banarjee}).


In sum, {the results presented here} are, to the best of our knowledge, the first examples of  spacetime structures explicitly emerging from the thermodynamic limit alone.
Remarkably, in our construction, we do not need any further assumptions other than those of standard 
thermodynamic fluctuation theory. In this sense, {we have provided here further evidence that}
spacetime descriptions appear to be a purely macroscopic effect, due to the thermodynamic limit.
Our results are completely formal and general, as they are based on the geometry of thermodynamic fluctuations, which does not depend on the details
of the microscopic description.
The central result of the present manuscript is that the maximally symmetric spacetimes 
emerge as equilibrium configurations in the thermodynamic sense.
This  suggests a direct relationship between the symmetries of spacetime and thermodynamic equilibrium. Indeed, one would heuristically expect that a system
at thermodynamic equilibrium would have the maximum possible number of symmetries. 
Although we have not proven the uniqueness of such solutions, the class of Hessian and Einstein metrics
is very restricted in dimensions equal or grater than four \cite{AmariCurvature,EinsteinMetrics,ShimaBook}. 
In fact, the uniqueness can be proved in two and three spacetime dimensions.

The thermodynamic origin of the spacetimes that we have described poses a number of questions and directions for future work. 
First of all, it will be interesting to find different spacetime solutions -- e.g. black holes -- along the same lines of reasoning. 
We believe that this can either be achieved by relaxing the requirement of full thermodynamic equilibrium or by using different statistics, such as
 R\'enyi or Tsallis' statistics \cite{Renyi, Tsallis}.
Secondly, 
 we have not investigated here the physical properties of the EGB {vacuum solution} that emerges as the unique geometry ruling thermodynamic fluctuations in the TPS.
We believe that a full physical understanding of this can also shed light over the physical significance of the embedded spacetimes. In this manner, our construction  resembles  a recent proposal to interpret spacetime within string theory based on the concept of Born reciprocity \cite{freidel}. 
In  third place, {the proportionality between the gravitational and the thermodynamic potentials,} the area law for the entropy and the direct {identification}
 between the thermodynamic potential and the classical action 
have appeared naturally in our framework, without the need
to impose any further assumptions. In our view this might have a profound significance.
Finally, we want to comment that although we have not proved that gravity itself is of thermodynamic origin -- in principle the gravitational interaction can also be present
in the microscopic theory -- nevertheless we have provided here for the first time a completely formal derivation of spacetimes solutions emerging just from the thermodynamic limit, thus
revealing new arguments in favor of the possible thermodynamic nature of space and time.  

The authors would like to thank F. Nettel for interesting discussions.
AB acknowledges the A. della Riccia Foundation (Florence, Italy) for financial support. CSLM is funded by a UNAM-DGAPA post-doctoral fellowship.
This work was supported by DGAPA-UNAM, Grant No. 113514, Conacyt-Mexico, Grant No. 166391.


\end{document}